\documentclass[a4paper,10pt]{article}
\usepackage[utf8]{inputenc}
\usepackage{ucs}
\usepackage[utf8]{inputenc}
\usepackage{amsmath}
\usepackage{amsfonts}
\usepackage{amssymb}
\usepackage{fontenc}
\usepackage{graphicx}
\usepackage{cite}

\title{Null second order corrections to Casimir energy in weak gravitational field: The Schwinger's approach}
\author{A.P.C.M. Lima$^a$\footnote{augustopcml@gmail.com}, G. Alencar$^a$ and R.R. Landim$^a$}
\date{{\it ${}^a$Departamento de F\'{\i}sica, Universidade Federal do Cear\'{a}-
Caixa Postal 6030, Campus do Pici, 60455-760, Fortaleza, Cear\'{a}, Brazil.}\\
\bigskip
\today}

\begin{document}

\maketitle

\begin{abstract}
We show through Schwinger's approach that in a static weak gravitational background, the Casimir Energy for a real massless scalar field obeying Dirichlet boundary conditions on rectangular plates is unaltered from it's flat space-time value. The result is obtained considering an rather general class of backgrounds, adding further generality and consistency test for the previous work. The proposed result has direct consequences on earlier works in the literature that found gravity related corrections for similar setups.
\end{abstract}

\section{Introduction}

The origin of effect dates back to 1948, with the first paper on the subject by H. Casimir\cite{casimir1948}. The situation considered is that of an electromagnetic vacuum ``trapped'' between parallel conducting plates. The presence of these plates forces boundary conditions on the fluctuations, perturbing the vacuum and generating an attractive force between the plates. It turns out that the mathematical essence of this phenomenon is quite general, and many more cases can be exploited by simply changing the characteristics of the field spin and the topology of boundary conditions. 

The Casimir effect is now a highly accepted phenomenon, with first accurate measurements emerging by the end of the 1990 decade, with \cite{lamoreaux,mohideen}, while more modern experimental progresses are also present, see for example \cite{Chang2011,Banishev:2012zz,Chang:2012fh,Banishev:2012fi,Banishev:2012bh}. An experimental discussion that is closely related to the contents of this manuscript is the discussion of vacuum weight experiments\cite{Avino:2019fdq}. 

From the various aspects and generalizations that can be brought to the table on the study of vacuum energy, our target will be the behavior of vacuum energy in a curved background space-time. This study has an possible cosmological interest, as an intrinsic space-time energy density could be related for example, to a cosmological constant (or non-constant). Also, it is of general theoretical interest to discuss how(or if) quantum fluctuations do gravitate. The study of Casimir effect in non-trivial background geometries has been already thoroughly approached in the literature (\cite{Bimonte:2008zva, dowker, Ford:1976fn,Altaie:1978dx,Baym:1993zx, Karim:2000ja, Setare:2001nx,Elizalde:2002dd,Aliev:1996va, Elizalde:2006iu, BezerradeMello:2006du, Elizalde:2009nt, NouriZonoz:2009zq, Nazari:2011hp, Saharian:2010ju, Bezerra:2011zz, Bezerra:2011nc, Milton:2011kz,BezerradeMello:2011nv, Sorge:2018zfd, Sorge, Nazari, stabile, Bezerra,Blasone:2018nfy,Buoninfante:2018bkc,Zhang,Bezerra:2016qof,Klimchitskaya:2014psa,Klimchitskaya:2015kxa, Sorge:2014uma, Muniz:2013uva, Quach:2015qwa} and many others).

More specifically, the focus of this work is calculating the energy shift due to to a weak static gravitational background, measured by a static observer. Our discussion is directed towards the original setup of \cite{Sorge}, this result was well received and has been used as basis for different models, examples being \cite{Nazari,Blasone:2018nfy,stabile,Bezerra,Buoninfante:2018bkc}. However, a critical turn from it is proposed in \cite{Lima:2019pbo}, where a mathematical inconsistency is shown in the original calculation, leading to the result that actually no energy shift is found to the order of$[M/R]^2$. In \cite{Sorge:2019ldb} Sorge replicates, reiterating his previous result, but is followed by another critique in \cite{Lima:2020egt}.

In this work we use the same formalism as \cite{Sorge:2019ldb,Lima:2020egt}, the Schwinger's approach\cite{SchwingerI,SchwingerII}, to reiterate and further generalize the result of \cite{Lima:2019pbo}. The background considered can cover a large class of background space-times, including those of previous works, as we only require that approximate planar symmetry is kept inside the small Casimir cavity. The result also adds consistency to the previous result and shows a more direct method of obtaining the vacuum energy. We remark the relevance of this critical turn as it call for revision of a few earlier results, and foments once again the discussion of how gravity can influence vacuum fluctuations.

\section{Definition of the problem}

In this section we provide a brief review of the setup involved. We take the initial considerations from \cite{Sorge}, and discuss the alternative Schwinger's method. In the context of curved space-times, we propose on how to relate the energy values in both formalisms.

The system in consideration is a pair of parallel rectangular plates with surface A, separation L and distance R from the origin of the radial coordinate system to the center of the closer plate. These plates are immersed in a weak gravitational background field
\begin{equation}
g_{\mu\nu}\simeq \eta_{\mu\nu}+h_{\mu\nu},
\end{equation}
 on which we will assume the perturbation to be static (independent of the timelike component and $h_{0i}=h_{ij}=0$ for $i\neq j$). For simplicity, the plates are assumed to be placed orthogonally to the radial direction and that $R>>\sqrt{A}>>L$. The measures are taken by an ``static observer'', i.e. with world velocity:
\begin{equation}\label{velocity}
u^\mu=\vert g_{00}\vert^{-1/2}\delta_0^\mu.
\end{equation}
The quantity we wish to find is the mean proper vacuum energy density of a real massless scalar field, defined as
\begin{equation}\label{energy1}
\bar{\epsilon}=\frac{1}{V_p} \int d^3x\sqrt{h} u^\nu u^\mu\langle 0 \vert T_{\mu\nu}\vert 0 \rangle.
\end{equation}

The expected value of the relevant energy-momentum tensor component can be calculated through standard methods, such as using the green functions for the field or considering the mode expansion as done in \cite{Sorge}. However, motivated by \cite{Sorge:2019ldb}, we use the alternative Schwinger's approach, which we briefly review next.

\subsection{Schwinger's approach}
Consider the vacuum persistence amplitude for large times in the absence of external sources:
\begin{equation}
_+\langle 0\vert 0 \rangle_-=Z[0]=e^{-iW[0]}.
\end{equation}
In flat space-time, the functional $W[0]$ can be interpreted as the vacuum energy density multiplied by a space-time volume factor $V^{(4)}=ALT$. Consider the action for the scalar field writen in the form
\begin{equation}
 S=\int dv_x[\phi(x)\hat{K}\phi(x)],
\end{equation}
where $\hat{K}$ is the wave operator (or commonly used proper-time hamiltonian $\hat{H}$ which coincides with $\hat{K}$ in the massless case) and $dv_x$ is the invariant volume element. It can be shown through an eigenfunction expansion
\begin{align}
&\hat{K}\phi_n(x)=\lambda_n\phi_n(x)\label{eigen1},\\ 
&\int dv_x \phi_n^*\phi_m=\delta_{nm}\label{normal1},
\end{align}
that $W[0]=i\ln(Z[0])$ reduces to
\begin{equation}
W[0]=\frac{i}{2}\int_0^{\infty} \frac{ds}{s}Tr[e^{-is\hat{K}}],
\end{equation}
where the trace is to be taken over all the eigenvalues defined in condensed notation in (\ref{eigen1})(or any equivalent representation). By using the orthonormality condition (\ref{normal1}), we can write the above equation as
\begin{equation}\label{w1}
 W[0]=\frac{i}{2}\sum_n\int dv_x\int \frac{ds}{s} \vert \phi_n(x)\vert^2 e^{-is\lambda_n}.
\end{equation}
This is the equivalent to the expression shown in \cite{Sorge:2019ldb}, using however formal eigenfunctions and eigenvalues of $\hat{K}$, which is, is our case, the curved space-time d'Lambertian.

Expression (\ref{w1}) is an invariant quantity related to the vacuum-vacuum transition rates of the system rest(comoving) frame (oberserver \ref{velocity}). In analogy to the flat case, we need to factor out the invariant volume space-factor to obtain (\ref{energy1})
\begin{equation}\label{energy2}
 \tilde{W}[0]=V_p^{(4)}\bar{\epsilon}.
\end{equation}
This will cause an divergence from the result obtained in \cite{Sorge:2019ldb}, which considers a non-invariant volume in the above expression.

\section{Casimir Energy for the real scalar field immersed in a generic static weak gravitational background}
\subsection{Background geometry}
Now we proceed to find the Casimir energy for the scalar vacuum in the weak gravitational background, but first we discuss the background to be considered. We begin with the metric used in \cite{Sorge}, which is the one obtained from the usual vacuum solution with spherical symmetry on weak field approximation 
\begin{equation}
 ds_{weak}^2=(1+2\Phi(r))dt^2+(1-2\Phi(r))dl^2,
\end{equation}
where $\Phi(r)=M/r$. The metric is then expanded in a rectangular system with radial direction being taken as the $z$ axis. Terms are kept up to second order in $M/R$, by considering its local form, we can get rid of the constant first order terms, leaving us with
\begin{equation}\label{metric1}
ds^2=(1-2\gamma z)dt^2+(1+2\gamma z)(dx^2+dy^2+dz^2). 
\end{equation}
A more general case is considered in \cite{Nazari}, on which generic space-times are considered, but retaining the spherical symmetry, which accounts to put the metric in the form
\begin{equation}\label{metric2}
\bar{ ds}^2=(1+2 \gamma_1 z)dt^2-(1-2\gamma_2 z)(dx^2+dy^2+dz^2).
\end{equation}
A few other works used an similar or equivalent form for the metric to calculate casimir energy corrections taking after \cite{Sorge}, such as \cite{stabile,Blasone:2018nfy,Buoninfante:2018bkc,Bezerra}.

For further completeness we want to introduce a more general metric, but we avoid one that would make the calculations too complicated. We will require then that the local metric keeps being only $z$ dependent(to order of $\gamma$) so that the line element is a constant on all planes parallel to the plates, but let each component have an distinct coefficient
\begin{equation}\label{metric3}
 \tilde{ds}^2=(1+2\gamma_1 z)dt^2-(1-2\gamma_2 z)dx^2-(1-2\gamma_3 z)dy^2-(1-2\gamma_4 z)dz^2,
\end{equation}
making this metric indeed more general than the previous ones. The requirement of planar symmetry is important here because the proper(measured) separation of the plates, which the final result
should expressed as an function of, needs to proportional to the coordinate distance L. Otherwise direct comparing this result to the flat case would lose physical sense. Alternatively, we could  impose that the measured distance is constant along the plates and work with a variable coordinate length, but would end up with very complicated expressions. 

Most known solutions of Einsteins equations assume highly symmetric distributions, so that weak field approximations in the form of (\ref{metric3}) that do not fall back in (\ref{metric2}) will seldom appear in literature. Nevertheless, it doesn't rule out the fact that they may describe physical situations. Also, imposing as few constraints as possible without escaping the present purposes helps ruling out the possibility of coincidental results, further attesting generality to the effect.   

\subsection{Eigenvalue equation}

Now we solve the eigenvalue equation for the operator $\hat{K}$, which in the present case is the curved space-time 
d'Lambertian
\begin{equation}\label{KG}
\hat{K}\phi_n=\vert g\vert^{-1/2}\partial_\mu[\vert g\vert^{1/2}g^{\mu\nu}\partial_\nu\phi_n]=\lambda_n\phi_n.
\end{equation}
For the metric (\ref{metric3}), this expression takes the form 
\begin{equation}\label{eigen2}
[(1-2\gamma_1z)\partial_t^2-(1+2\gamma_2z)\partial_x^2-(1+2\gamma_3z)\partial_y^2-(1+2\gamma_4z)\partial_z^2-\gamma_5\partial_z]\phi_n=\lambda_n\phi_n, 
\end{equation}
where $\gamma_5=\gamma_1-\gamma_2-\gamma_3+\gamma_4$. A much similar equation is solved in \cite{Sorge}, so we will follow in close analogy to it. First expand the eigenfunctions as:
\begin{equation}
\phi_n(x)=A_n\chi_n(z)e^{i\omega_n t-k_\perp x_\perp}.
\end{equation}
We will keep the simplified $n$ index notation, but it is now to be understood that the eigenvalues have four degrees of freedom. With that,  (\ref{eigen2}) becomes
\begin{equation}
 [(1-2\gamma_1z)\omega^2-(1+2\gamma_2z)k_x^2-(1+2\gamma_3z)k_y^2+(1+2\gamma_4z)\partial_z^2+\gamma_5\partial_z]\chi=-\lambda_n\chi, 
\end{equation}
this can be rearranged into
\begin{equation}\label{eigen3}
 \chi''+\gamma_5\chi'-az\chi+b\chi=0,
\end{equation}
where
\begin{align}
&a=2\gamma_1\omega^2+2\gamma_2k_x^2+2\gamma_3k_y^3+2\gamma_4b,\label{a}\\
&b=\omega^2-k_x^2-k_y^2+\lambda.\label{b}
\end{align}
The first order derivative term can be simplified using
\begin{equation}
\chi=e^{-\gamma_5z/2}\varPsi,
\end{equation}
so (\ref{eigen3}) becomes 
\begin{equation}
\varPsi''-az\varPsi+b\varPsi=0 
\end{equation}
which in turn, through the transformation
\begin{equation}
u(z)=-a^{1/3}z+ba^{-2/3}, 
\end{equation}
becomes the Airy differential equation
\begin{equation}
\varPsi''+u\varPsi=0. 
\end{equation}
The resulting solution expressed in terms of Bessel functions, which can in turn be expanded in asymptotic form to order of $\gamma$, becoming
\begin{equation}
 \varPsi=\kappa u^{-1/4}\sin\left(\frac{2}{3}u^{3/3}+\varphi\right),
\end{equation}
where $\kappa$ e $\varphi$ are constants. Then the full solution reads
\begin{equation}
 \phi_{\omega,k_\perp,n}(x)=\kappa\sin\left(\frac{2}{3}u^{3/3}+\varphi\right)e^{i(\omega t-k_\perp x_\perp)-\gamma_5 z/2}.
\end{equation}

Applying the Dirichlet boundary conditions $\phi(z=0)=\phi(z=L)=0$, results in
\begin{equation}
u(L)-u(0)\simeq n\pi, 
\end{equation}
that to first order in $\gamma$ reads 
\begin{equation}\label{misc1}
 b-aL/2=\tilde{n}^2
\end{equation}
where $\tilde{n}=n\pi/L$. Using (\ref{a},\ref{b}) and (\ref{misc1}), we can finally find the eigenvalues
\begin{equation}\label{eigenvalues}
 \lambda=\tilde{n}^2(1+\gamma_4 L)+k_x^2(1+\gamma_2 L)+k_y^2(1+\gamma_3L)-\omega^2(1-\gamma_1L),
\end{equation}
these can be seen as an remnant of the dispersion relation.

\subsection{Casimir energy density}

With the values (\ref{eigenvalues}) in hand we can proceed to a straightforward calculation of the Casimir energy. Plugging those back in (\ref{w1}) and using the regularization factor from \cite{farina}, we get
\begin{align}
W^{(\nu)}[0]=\frac{i}{2}\sum_n\int d^4x \sqrt{-g}\int d^2k_\perp d\omega \int ds s^{\nu-1}\vert \phi_{\omega,k_\perp,n}\vert^2\nonumber\\
\times e^{-is[\tilde{n}^2(1+\gamma_4 L)+k_x^2(1+\gamma_2 L)+k_y^2(1+\gamma_3L)-\omega^2(1-\gamma_1L)]}.
\end{align}

The calculation carries in a much similar way to the flat space-time case, except for the extra factors coming from the $s$ exponential. Integration in $k_\perp,\omega$ and space-time yields
\begin{align}
&W^{(\nu)}[0]=[1+(-\gamma_1+\gamma_2+\gamma_3)L]^{-1/2}\frac{\sqrt{i}AT}{2(2\pi)^3}\nonumber\\
&\times\sum_n\int ds s^{\nu-1}\exp\left[\left(-is\frac{n^2}{\pi^2 L^2}\right)(1+\gamma_4 L)\right]\left(\frac{\pi}{is}\right)^{3/2}\nonumber\\
&=[1+(-\gamma_1+\gamma_2+\gamma_3)L]^{-1/2}\frac{\sqrt{i} A T}{16\pi^{3/2}}\Gamma(\nu-3/2)\sum_n\left[\left(i\frac{n^2\pi^2}{L^2}\right)(1+\gamma_4 L)\right]^{3/2-\nu}\nonumber\\
&=[1+(-\gamma_1+\gamma_2+\gamma_3)L]^{-1/2}(1+\gamma_4 L)^{3/2-\nu}\frac{\pi^{3/2} A T}{16 L^{3-2\nu}}\Gamma(\nu-3/2)\zeta(2\nu-3),
\end{align}
where an analytical continuation to the zeta function is to be understood in the last equality. Restoring the original value $\nu=0$ gives the renormalized $W^{(0)}[0]$
\begin{equation}\label{w}
 W^{(0)}[0]=-\left[1+\frac{L}{2}(\gamma_1-\gamma_2-\gamma_3+3\gamma_4)\right]\frac{AT\pi^2}{1440 L^3}.
\end{equation}

Now, there are two steps left, we need to use (\ref{energy2}) to get $\bar{\epsilon}$ from (\ref{w}) and express the proper energy density in terms of invariant quantities. The proper space-time volume is given by
\begin{equation}
 V_p^{(4)}=\int d^4x \sqrt{-g}\simeq ALT\left[1+\frac{L}{2}(\gamma_1-\gamma_2-\gamma_3-\gamma_4)\right],
\end{equation}
thus
\begin{equation}\label{energy3}
\bar{\epsilon}=\frac{W^{(0)}[0]}{V_p^{(4)}}=-(1+2\gamma_4 L)\frac{\pi^2}{1440 L^4}.
\end{equation}
The coordinate parameter $L$ should finally be expressed in terms of the proper separation $L_p$ given by
\begin{equation}
 L_p=\int dz \sqrt{-g_{33}}\simeq L(1-\gamma_4 L/2).
\end{equation}
On inverting the above relation and plugging back into (\ref{energy3}) we finally get
\begin{equation}
\bar{\epsilon}=-\frac{\pi^2}{1440 L_p^4}. 
\end{equation}
This extends the result from \cite{Lima:2019pbo} to the metric (\ref{metric3}).

\section{Concordance with the mode expansion method}

The result obtained in the last section is in agreement with the ones from \cite{Lima:2019pbo}, however, as a remark regarding the mentioned work is made in \cite{Sorge:2019ldb} that could directly affect the result, we would like to address it here. 

The specific claim is that the mode solution presented in \cite{Lima:2019pbo} does not satisfy proper orthonormalization conditions. As no explicit demonstration is made, we would like to present ours, and reinforce the previous result, to which the method used in this paper shows agreement in a more general class of space-time metrics(\ref{metric3}). 

We go back to the simplest case of space-time (\ref{metric1}). The normalization used in the canonical approach was
\begin{equation}\label{prod}
\langle \phi_m, \phi_n\rangle =i\int d\Sigma\sqrt{|h|}u^{\mu}(\psi_n^*\partial_\mu\psi_m-\psi_m\partial_\mu\psi_n^*) =\delta^2(k_m-k_n)\delta_{n,m}. 
\end{equation}
Where $\Sigma$ is a timelike hypersurface we take as $t=0$ and $u^\mu$ its normal outward vector(which can be identified with the four velocity (\ref{velocity})). The modes considered are
\begin{equation}\label{solutions}
 \phi_n(x)=\left[\frac{1}{2\pi\sqrt{\omega_{0,n}}}\sin(n\pi z/L)+\gamma\chi^{(\gamma)}(z)\right]e^{i(\omega_n t-k_\perp x_\perp)}=\chi_n(z)e^{i(\omega_n t-k_{\perp} x_\perp)},
\end{equation}
where
\begin{align}
&\omega_{0,n}=\sqrt{\frac{n^2\pi^2}{L^2}+k_\perp^2},\\
&\omega_n\simeq (1+\gamma L)\omega_0,
\end{align}
and
\begin{equation}
\chi^{(\gamma)}=\frac{2n\pi\omega_0^2L^2(L-z)\cos(n\pi z/L)+L(2n^2\pi^2z+2k^2L^2z-k^2L^3)\sin(n\pi z/L)}{4Ln^2\pi^3\sqrt{\omega_0 L}}.
\end{equation}

Equation (\ref{prod}) for these modes can be written as
\begin{equation}
\langle \phi_m, \phi_n\rangle=(\omega_n+\omega_m)\int d^3x (1-4\gamma z)\chi_n\chi_m e^{i(k_m-k_n)},
\end{equation}
which can be checked directly with the use of a symbolic integration software or some extensive algebra. The transversal $(x,y)$ parts of the integral already trivially yield delta functions(factoring a $4\pi^2$ constant) while the remaining $z$ integral results in
\begin{align}
 &4\pi^2(\omega_n+\omega_m)\int_0^L dz (1-4\gamma z)[\chi_n(z)\chi_m(z)]=(1-2\gamma L)\frac{\omega_n+\omega_m}{\sqrt{\omega_{n,0}\omega_{m,0}}}\delta_{n,m}+O[\gamma^2]\nonumber\\
&=\delta_{n,m}+O[\gamma]^2.
\end{align}
Thus, condition (\ref{prod}) is satisfied to relevant order. Moreover, the above solutions also satisfy the $\lambda=0$ case of (\ref{KG}), i.e. the field equation. Then , to the best of this analysis no problem is shown with the presented solution.

\section{Conclusion}

We have worked out the Casimir energy density for a massless real scalar field obeying Dirichlet conditions on parallel plates in a weak static gravitational background (\ref{metric3}). The metric used here covers a wider class of space-time configurations than similar previous works such as \cite{Nazari} and it was shown that there is apparently no energy shift associated with gravity (at least to first nontrivial order, i.e. second order) in those cases, contrary to the results of \cite{Sorge,Sorge:2019ldb}. The result presented here adds further generality to the ones from \cite{Lima:2019pbo,Lima:2020egt}, where the vacuum energy is calculated using a mode expansion method. Moreover, it also serves as further consistency test by obtaining the same conclusion as \cite{Lima:2019pbo} while following a much different approach.

As argued before, a few other factors might have to be taken in consideration for a more general and consistent analysis, such as for example, plate finiteness or considering different vacuum states. However,as far as to the prior considerations made in \cite{Sorge}, and the following analysis from some works \cite{Blasone:2018nfy,stabile, Bezerra,Buoninfante:2018bkc} that take it as a starting point, the lack of an energy shift pointed here is of critical importance, as it may drastically change the nature of their results.  

Although the absence of a gravitational correction to the Casimir energy in the present case might seen odd at first, we believe further analysis might shed some clues on it. Comparing this result with strong and non-static regimens might provide us with further insight on the study of vacuum energy.

\section*{Acknowledgements}

The authors would like to thank Alexandra Elbakyan and sci-hub, for removing all barriers in the way of science.

We acknowledge the financial support  by Conselho Nacional de Desenvolvimento Cient\'ifico  e Tecnol\'ogico(CNPq) and Funda\c{c}\~ao Cearense de Apoio ao Desenvolvimento Cient\'ifico e Tecnol\'ogico(FUNCAP) through PRONEM PNE0112-00085.01.00/16.

\end{document}